\documentclass[preprint,aip,superscriptaddress,preprintnumbers,amsmath,amssymb]{revtex4-1}
\usepackage{graphicx}
\usepackage[version=2]{mhchem}
\usepackage{subcaption}
\usepackage{array}
\usepackage{siunitx}
\usepackage{booktabs}
\usepackage{gensymb}
\usepackage{multirow}
\usepackage{dcolumn} 
\usepackage{setspace}
\usepackage{array}
\usepackage{psfrag}
\usepackage{textcomp}
\usepackage{color}
\begin{document}

\newcommand{\tny}[1]{\mbox{\tiny $#1$}}
\newcommand{\eqn}[1]{\mbox{Eq.\hspace{1pt}(\ref{#1})}}
\newcommand{\eqs}[2]{\mbox{Eq.\hspace{1pt}(\ref{#1}--\ref{#2})}}
\newcommand{\eqsu}[2]{\mbox{Eqs.\hspace{1pt}(\ref{#1},\ref{#2})}}
\newcommand{\eqtn}[2]{\begin{equation} \label{#1} #2 \end{equation}}
\newcommand{\func}[1]{#1 \left[ \rho \right] }
\newcommand{\mfunc}[2]{#1_{#2} \left[ \rho \right] }
\newcommand{\mmfunc}[3]{#1_{#2} \left[ #3 \right] }
\newcommand{\mmmfunc}[4]{#1_{#2}^{#3} \left[ #4 \right] }
\newcommand{\pot}[1]{v_{\rm #1}}
\newcommand{\spot}[2]{v_{\rm #1}^{#2}}
\newcommand{\code}[1]{\texttt{#1}}
\newcommand{\centerit}[1]{{ \begin{center} #1 \end{center} }}
\def\brpppp{{\mathbf{r}^{\prime\prime\prime\prime}}}
\def\brppp{{\mathbf{r}^{\prime\prime\prime}}}
\def\brpp{{\mathbf{r}^{\prime\prime}}}
\def\brp{{\mathbf{r}^{\prime}}}
\def\bzp{{\mathbf{z}^{\prime}}}
\def\bxp{{\mathbf{x}^{\prime}}}
\def\tp{{{t}^{\prime}}}
\def\tpp{{{t}^{\prime\prime}}}
\def\tppp{{{t}^{\prime\prime\prime}}}

\def\tbr{{\tilde{\mathbf{r}}}}
\def\bk{{\mathbf{k}}}
\def\br{{\mathbf{r}}}
\def\bz{{\mathbf{z}}}
\def\bx{{\mathbf{x}}}
\def\bR{{\mathbf{R}}}
\def\bM{{\mathbf{M}}}
\def\bP{{\mathbf{P}}}
\def\bT{{\mathbf{T}}}
\def\bK{{\mathbf{K}}}
\def\bA{{\mathbf{A}}}
\def\bB{{\mathbf{B}}}
\def\bX{{\mathbf{X}}}
\def\bY{{\mathbf{Y}}}
\def\bP{{\mathbf{P}}}
\def\bI{{\mathbf{I}}}
\def\d{{\mathrm{d}}}
\def\rhor{{\rho({\bf r})}}
\def\rhorp{{\rho({\bf r}^{\prime})}}
\def\rhoi{{\rho_I}}
\def\rhoii{{\rho_{II}}}
\def\rhoj{{\rho_J}}
\def\rhoir{{\rho_I({\bf r})}}
\def\rhoiir{{\rho_{II}({\bf r})}}
\def\rhojr{{\rho_J({\bf r})}}
\def\rhoirp{{\rho_I({\bf r}^{\prime})}}
\def\rhojrp{{\rho_J({\bf r}^{\prime})}}
\def\sumi{{\sum_I^{N_S}}}
\def\sumj{{\sum_J^{N_S}}}
\def\im{{\operatorname{Im}}}
\def\dscf{{$\Delta$SCF}}

\def\etal{{\it et al.}}
\def\vdw{{van der Waals}}
\def\qe{{\sc Quantum ESPRESSO}}
\def\se{{Schr\"{o}dinger equation}}
\def\ses{{Schr\"{o}dinger equations}}
\def\bnabla{{\boldsymbol{\nabla}}}
\def\bchi{{\boldsymbol\chi}}
\def\bLambda{{\boldsymbol\Lambda}}
\def\bDelta{{\boldsymbol\Delta}}

\def\Lcal{{\ensuremath{\mathcal{L}}}}
\def\Dcal{{\ensuremath{\mathcal{D}}}}
\def\Pcal{{\ensuremath{\mathcal{P}}}}
\title{Static Correlation Density Functional Theory}

\author{Pablo Ramos} 

\affiliation{Department of Chemistry, Rutgers University, Newark, NJ 07102, USA}

\author{Michele Pavanello}
\email{m.pavanello@rutgers.edu}

\affiliation{Department of Chemistry, Rutgers University, Newark, NJ 07102, USA}
\affiliation{Department of Physics, Rutgers University, Newark, NJ 07102, USA}

\begin{abstract}
Over the years, several schemes have been proposed to describe multireference systems with Kohn-Sham Density Functional Theory. Problematic is the combination of two aspects: the Kohn-Sham reference wavefunction is usually taken to be a single Slater determinant, and approximate exchange-correlation functionals are typically derived form the local density approximation. In this work, we develop a theoretical framework that foregoes the single Slater determinant and instead employs thermal states as reference states for zero-temperature interacting systems. We provide convenient definitions of static and dynamic correlation functionals via an adiabatic connection approach. The formalism and computational results indicate that the entropic term of the thermal reference state is a good approximation to the static correlation functional. Hence, this work validates several reported results in the literature and motivates additional developments of static correlation density functionals.
\end{abstract}

\date{\today}
\maketitle

\section{Introduction}
The Kohn-Sham formulation of Density Functional Theory (KS-DFT) \cite{kohn1965} provides the possibility of solving for the electronic structure of a set of noninteracting electrons (the Kohn-Sham system) in place of the real, interacting system. The KS system is subject to an external potential, $v_s(\br)$, that differs from the one of the real system in such a way that the KS and real electron densities match.  The real systems for which this procedure can be carried out successfully are called \emph{noninteracting $v_s(\br)$ representable} \cite{Schipper_1998,Gritsenko_1997}. 

Over the decades, several studies have been carried out to characterize the $v_s$ representability of given (physical or not) electron densities \cite{Lieb_1983}. An important finding was that when the real system is represented by a multireference wavefunction, the associated KS system is better represented by an ensemble \cite{Schipper_1998}. It is important to point out that an ensemble of noninteracting electrons is still a valid KS system \cite{Savin_2003,Englisch_1984}. Thus, several studies have emerged exploiting ensemble states as more convenient KS references than pure states \cite{Wang_1996} particularly for those cases where degeneracy arises (e.g., bond forming and breaking, atomic systems). 

Historically, the first occurrence of the use of fractional occupations for mean-field calculations of molecules and materials is due to Slater in Ref.\citenum{Slater_1969} where his team applied fractional occupations to solutions of Hartree-Fock equations as well as the $X^\alpha$ method. More recently, the use of thermal ensemble states (thermal states, hereafter) has been advocated by several groups in several different contexts. Grimme showed that the energetics and electronic structure of fragmented molecules is better reproduced by thermal KS states when effective temperatures of several thousand Kelvin are adopted \cite{Grimme_2013}. They also showed that the density difference between approximate KS-DFT calculations (i.e., employing approximate exchange-correlation (xc) functionals) of thermal states and pure states is a useful measure of the static correlation character of a molecular system \cite{Grimme_2015}. Chai has showed that when the electronic entropy of the ensemble KS system is included in the evaluation of the electronic energy, potential energy curves of bond dissociation \cite{Chai_2012}, transition states of chemical reactions \cite{Chai_tao_2017} and other properties \cite{Chai_tao_2019, Chai_tao_2018, Chai_tao_2018_2, Chai_tao_2017_2, Chai_tao_2017_3, Chai_tao_2016, Chai_tao_2016_2} are dramatically improved compared to the standard use of pure states and approximate xc functionals. Thermal states are commonplace when applying KS-DFT to metallic systems where without the use of effective temperatures the nonlinear self-consistent field procedure would not converge \cite{Gillan_1989,Slater_1969}.

Another effective way to deal with near degeneracies when they appear in either the spectrum of the real Hamiltonian or the KS Hamiltonian is to break spin symmetry or the spatial point-group symmetry. For example, it is known that spin unrestricted calculations of bond dissociation lead to qualitatively correct bond breaking energy curves at the expense of generating completely unphysical electron densities. 
 
The term static correlation (also known as nondynamic correlation) is commonly adopted to characterize the errors that arise when approaching multireference systems with single Slater determinant wavefunctions. These errors are purely the result of correlation effects both in the correlated wavefunction world as well as in the DFT world. It is well understood that currently available xc functionals do not correctly capture static correlation effects \cite{cohen2008a}. So-called flat-plane conditions \cite{Cohen_2011} have been derived linking the ability of approximate xc functionals to approach multireference ground states to their ability to predict degenerate energies when evaluated on the electron densities associated with each of the configuration state functions. These exact conditions guide future xc density functional developments. Thus far, however, pure xc density functionals still generally \cite{Becke_2003,Becke_2005} lack the ability to capture static correlation.

The static correlation problem of mean-field methods has been an active topic of research for decades with reduced density matrix functional theory \cite{Muller_1984,Buijse_2002,Kamil_2016,Lathiotakis_2009,Hellgren_2019} and range separatated DFT \cite{Stoyanova_2013,Sharkas_2012,Fromager_2015} being perhaps the most active lines of work.

In this work, we start from the following ansatz of the correlation energy \cite{Cioslowski_1991}
\eqtn{in1}{E_c[\rho]=E_c^\text{static}[\rho]+E_c^\text{dynamic}[\rho],}
which we approach by defining $E_c^\text{static}[\rho]$ first, and then $E_c^\text{dynamic}[\rho]=E_c[\rho]-E_c^\text{static}[\rho]$. $E_c^\text{static}[\rho]$ needs to be defined appropriately so that the commonly employed approximate xc density functionals can still be considered legitimate approximations to the dynamic correlation, $E_c^\text{dynamic}[\rho]$. 

This is a point that differentiates the approach we employ here with range separation. In range separation, the strong (static) correlation is identified as a long range effect \cite{Fromager_2015}, while in this work the definition of static correlation is given up front and dynamic correlation follows as the remainder. Clearly the word choice ``static'' and ``dynamic'' are somewhat arbitrary in this context. However, the aim is to include in static correlation only the component of the correlation energy that comes in when degeneracy is present.

The separation of correlation into static and dynamic has been pursued for many years \cite{Ludena_1997,Matito_2016, Cioslowski_1991} and generally it can be achieved once the full quantum problem has been solved. That is, once the Full CI (FCI) solution of the electronic problem in some appropriate one-electron basis set is achieved, it is possible to devise several (more or less chemically or physically insightful) paths connecting a mean-field solution (such a Hartree-Fock) to the FCI solution.

In this work, we take a different approach that adheres to these guiding principles: (1) Smearing the occupation of frontier orbitals according to the Fermi-Dirac distribution generates quality ensemble KS states that improve upon the single Slater determinant KS reference \cite{Grimme_2013,Grimme_2015}; (2) Static correlation is intimately related to (near)degeneracy, and thus it can be related to the fractional occupation of the frontier KS orbitals \cite{Savin_2000,Gersdorf_1997}; (3) The electronic entropic energy term is a good approximation to a static correlation energy functional \cite{Chai_2012,Gersdorf_1997}. We will show that these guiding principles result in a rigorous framework, involving an adiabatic connection approach for the correlation energy. We conclude the paper by discussing several computational examples on bond breaking potential energy curves. 

\section{Thermal Nonintercting KS Reference Systems}
In this section we provide two independent justifications for employing noninteracting thermal states as KS reference. The first is of practical nature; it shows that in a regime of strong static correlation the free energy of  thermal KS states computed with approximate (semilocal) xc functionals is better than the one from conventional semilocal KS-DFT computed with the same approximate xc functional. The second is based on a relation between the correlation energy and an adiabatic connection integral involving canonical one-body reduced density matrix (1-RDM) diagonal occupations.
Our analysis results in a static correlation functional that, even in its first formulation improves upon current xc functional approximants and paves the way to search for optimal static correlation functionals.

\subsection{Thermal states as noninteracting reference}


Let us consider an approximate xc functional, $\widetilde{E}_{xc}[\rho]$, and the exact functional, $E_{xc}[\rho]$. We break down the energy into various contributions for the thermal reference and for the single Slater determinant KS reference states. We construct the energy of the thermal state in such a way to feature an approximate $\widetilde{E}_{xc}[\rho]$ as well as an additional entropic term, $S_{xc}$. This term is introduced to correct for the approximate xc functional. The energies of the two references must be equal to each other as well as to the true total electronic energy,
\eqtn{th1}{T_s^\tau + \underbrace{E_H +\widetilde{E}_{xc}}_{\widetilde{W}} + V - \tau (S_s^\tau + S_{xc}) = T_s + \underbrace{E_H + {E}_{xc}}_{W} + V,}
where we have denoted $T_s^\tau$ as the thermal noninteracting kinetic energy, $S_s^\tau$ as the Shannon entropy of the thermal noninteracting system, and $W$ and $\widetilde{W}$ are the exact and approximate electron-electron interaction, respectively. Consequently, we define $S_{xc}$ as 
\eqtn{th3}{S_{xc}=\frac{1}{\tau}\bigg( \widetilde{W}-W + T_s^\tau - T \bigg)-S_s^\tau.}
It is clear that, given a value for the temperature $\tau$, there is a unique way of defining $S_{xc}$. 

In practical calculations, in a regime of strong static correlation (e.g., bond breaking), $\widetilde{E}_{xc}[\rho]$ is unable to reproduce correct energy and electron density. Specifically, because of the strong ionic character of the single Slater determinant wavefunction, the value of the approximate electron-electron interaction, $\widetilde{W}$, is much larger than the exact one, $W$.  Thus, even in the most extreme of the approximations (i.e., $S_{xc}=0$ which we will adopt later) thermal reference states should provide a more advantageous reference compared to the KS system precisely because the Shannon entropy, $S_s^\tau$, is a nonnegative quantity. We should also observe that the inequalities $T_s^\tau\geq T_s$ and $T\geq T_s$ typically hold, suggesting that the $T-T_s^\tau$ term is likely to be smaller than the commonly adopted correlation kinetic energy term, $T-T_s$.

Thus, from a practical perspective, the choice of using thermal noninteracting reference systems is now clear: the difference between the noninteracting and the interacting kinetic energy is smaller, and we can exploit error cancellation between the values of $S_s^\tau$ and the electron-electron interaction deviation, $\widetilde{W}-W$. 

\subsection{The adiabatic connection}
In this section we develop a framework that accomplishes \eqn{in1} in a convenient and straight forward way, finding $-\tau S_s^\tau$ to be a viable approximation to the static correlation functional.

Correlation energy can be defined using an adiabatic connection approach. This requires the integration of an appropriate kernel along a path where the potential for the electronic system is interpolated between the potential of a non-interacting system to the one of the fully interacting system. This approach can feature kernels of various kinds, such as derived from the electron-electron interaction or from the electronic kinetic energy \cite{Teale_2015}.  

Following Refs.\citenum{Savin_1995,Savin_2000}, it is possible to connect a potential describing a noninteracting system, $\hat{V}_0$ (such as the KS-DFT exact, $\hat{V}_0=\sum_i^{N_e} v_s(\br_i)$, with $N_e$ being the total number of electrons, or a KS-DFT approximate, $\hat{V}_0=\sum_i^{N_e} \tilde v_s(\br_i)$), with the true potential of the interacting system, $\hat{V}_1$, via a linear \cite{Harris_1974} interpolation scheme \footnote{For sake of simplicity, in \eqn{ac0} we adopt a linear interpolation for the potential. However, in Ref.\citenum{Savin_2000} the formalism does not require the interpolation to be linear.},
\eqtn{ac0}{\hat{V}_\lambda = (1-\lambda) \hat{V}_0 + \lambda \hat{V}_1.}
At this stage, we define a Hamiltonian, $\hat H_\lambda=\hat T + \hat V_\lambda$, and the ground state wavefunction resulting from $\hat H_\lambda \Psi_\lambda = E^\lambda\Psi_\lambda$. Thus, we can define the correlation energy at a particular coupling strength, $\lambda$, as
\eqtn{ac-i}{E_c^\lambda=E^\lambda-E^\lambda_0,}
where $E^\lambda_0=\langle \Psi_0 | \hat H_\lambda | \Psi_0 \rangle$.
The correlation energy can be written in terms of the following adiabatic connection integral
\eqtn{ac-4}{E_c = \int_0^1 \frac{\d E_c^\lambda}{\d\lambda} \d\lambda.}
Aided by the Hellmann-Feynman theorem, because $\frac{\partial V_\lambda}{\partial\lambda}=\hat V_1 -\hat V_0$, the above equation leads to
\eqtn{ac-2}{E_c=\int_0^1 \left( \langle \Psi_\lambda | \hat{V}_1 - \hat{V}_0 | \Psi_\lambda \rangle - \langle \Psi_0 | \hat{V}_1 - \hat{V}_0 | \Psi_0 \rangle \right)\d\lambda.}
The integrand can be refactored as
\eqtn{ac-3}{E_c = \int_0^1 \frac{\d \left(E_c^\lambda/\lambda\right)}{\d\lambda} \d\lambda,} 
which in view of \eqn{ac0}, combined with the observations that $\frac{\hat H_\lambda}{\lambda}=\frac{\hat T}{\lambda} + \frac{\hat V_0}{\lambda} + \text{constant}$, and that $\lim_{\lambda\to 0} \frac{E_c^\lambda}{\lambda}=0$\cite{Savin_2000}, leads to the equivalency of \eqs{ac-4}{ac-3}. 
Thus, the correlation energy becomes,
\eqtn{ac-1}{E_c=\int_0^1 -\frac{1}{\lambda^2} \left( \langle \Psi_\lambda | \hat{T} + \hat{V}_0 | \Psi_\lambda \rangle - \langle \Psi_0 | \hat{T} + \hat{V}_0 | \Psi_0 \rangle \right)\d\lambda.}
Unless it is chosen $\hat V_0=\sum_i v_s(\br_i)$, the adiabatic connection in the above equation admits an electron density, $\rho_\lambda(\br)=\langle \Psi_\lambda | \hat{\rho}(\br) | \Psi_\lambda \rangle$, that is not constant along the path $\lambda: 0\to 1$. 

The Hamiltonian $\hat H_0 = \hat{T} + \hat{V}_0=\sum_i \hat h_0 (\br_i)$ can be spectrally decomposed to $\hat h_0 = \sum_i \left| i \rangle \varepsilon_i \langle i \right|$. Thus, expressing the 1-RDM associated with $\Psi_\lambda$ in the canonical KS orbitals, $\{|i\rangle\}$, the correlation energy becomes
\eqtn{ac1}{E_c=\sum_i\varepsilon_i P_{ii},}
where 
\eqtn{ac2}{P_{ii}=\int_0^1 -\frac{1}{\lambda^2} d\lambda \left( P_i^\lambda - P_i^0  \right).}
$P_i^\lambda$ are the diagonal elements of the canonical 1-RDM at coupling strength $\lambda$ expressed on the basis of the canonical KS orbitals.

Let us reiterate that the above equations differ in spirit from the original formulation of the adiabatic connection \cite{lang1975,gunn1976} in that the electron density may not be kept constant along the $\lambda$ integration. This is important and allows us to pick as the noninteracting system one with $\hat{V}_0=\sum_i^{N_e} \tilde v_s(\br_i)$. Thus, a legitimate reference KS system is one computed with an approximate $\widetilde{E}_{xc}[\rho]$.

\subsection{Splitting the adiabatic connection integral into static and dynamic correlation paths}
\begin{figure}
\includegraphics[width=0.4\textwidth]{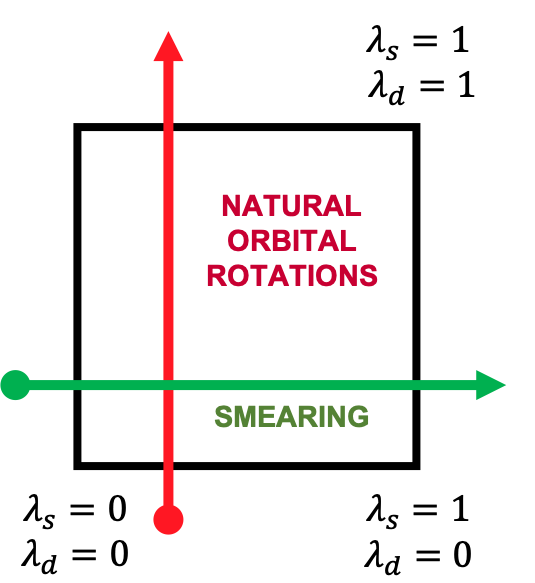}
\caption{\label{f_ac} A scheme depicting the proposed coupling integrations. The first coupling integration over $\lambda_s$ (green) is artificial and takes place over the horizontal path, only allowing ad-hoc smearing of the occupation numbers. The second integration over $\lambda_d$ (red), carries out the true coupling integration while phasing out the artificial integrand introduced in the first step. As $\lambda_d$ becomes greater than zero, the natural orbitals of the system will ``rotate away'' from the canonical KS orbitals.}
\end{figure}

The integral \eqn{ac2} can be split into two parts which we will associate with two different coupling constants (see Figure \ref{f_ac}):
\begin{itemize}
\item[$\lambda_s$] A \emph{static correlation} integral where the occupations of the canonical KS orbitals are varied according to an ad-hoc prescription. We indicate by $P_i^{s,\lambda}$ the occupations along this fictitious adiabatic connection path.
\item[$\lambda_d$] A \emph{dynamic correlation} integral where now (1) the natural orbitals of the interacting system at coupling strangth $\lambda_d$ ``rotate away'' from the canonical KS orbitals, (2) the diagonal occupations of the 1-RDM expressed in the basis of the canonical KS orbitals will follow the solution of the \se\ at the given coupling strength.  And, (3) at the same time, the ad-hoc prescription for the occupations carried out in with $\lambda_s$ is phased out.
\end{itemize}

Thus, the first, static correlation adiabatic connection integral over $\lambda_s$ will simply be
\eqtn{ac3}{P^\text{static}_{ii}=\int_0^1 -\frac{1}{\lambda_s^2} d\lambda_s \left( P_i^{s,\lambda_s} - P_i^0  \right),}
where $P_i^{s,\lambda_s}$ are the occupation numbers created by the smearing procedure. 
The second, dynamic correlation integral over $\lambda_d$ becomes, 
\eqtn{ac4}{P^\text{dynamic}_{ii}=\int_0^1 -\frac{1}{\lambda_d^2} d\lambda_d \left( P_i^{\lambda_d} - P_i^{s,\lambda_d}  \right).}

The ad-hoc prescription to generate the occupations along the $\lambda_s$ path should be chosen in a way that $P_i^{s,\lambda_s}$ is close to $P_i^{\lambda_d}$. Along a bond breaking reaction coordinate, this is achieved by any technique that will populate orbitals that are nearly degenerate. For example, the Fermi-Dirac distribution.

Splitting the adiabatic connection integral in two parts in \eqs{ac3}{ac4} allows us to split the correlation energy into two components as outlined by \eqn{in1}. An interesting observation is that in the absence of static correlation (no degeneracy and large gaps), $P_i^{s,\lambda_s} = P_i^0$ and therefore the static correlation integrand in \eqn{ac3} becomes identically zero and also $E_c^{\text{static}}=0$. Thus, as expected, the entirety of the correlation energy is dynamic, $E_c=E_c^{\text{dynamic}}$. This reinforces the common knowledge that the currently developed density functional approximants are good approximations to dynamic correlation.

\subsection{Change of variable, $\lambda\to P$, for the adiabatic connection integral}

There are two ingredients to the adiabatic connection integral in \eqn{ac3}. One involves formulating the smearing distribution function (here below we will choose Fermi-Dirac), and the other involves choosing a path connecting the Aufbau solution to the statically correlated one.  In regards to the latter, because the occupations are monotonic functions of the coupling strength \cite{Teale_2015}, $\lambda$, the integral \eqn{ac3} can be expressed in terms of an integration over the value of the occupations, from $P_i^0$ to $P_i^{s,1}$. Using a short-hand notation (i.e., dropping the subscript $i$ and the superscripts $s$ and $\lambda_s$) we have
\eqtn{ac6}{P_{ii}^\text{static}=\int_{P^0}^{P^1}f(P)(P-P^0)\d P,}
where the Jacobian associated with the transformation is 
\eqtn{ac7}{f(P)=\frac{\d \left( \frac{1}{\lambda_s} \right)}{\d P}.}
Thus, the formulation of an appropriate static correlation path connecting the smeared thermal state with the pure KS state translates to the formulation of an appropriate $f(P)$.

The function $f(P)$ should satisfy some physically imposed boundary conditions. 
\begin{itemize}
\item $f(P<0.5)>0$. The orbitals that are initially unoccupied, upon increasing of $\lambda_s$ will steadily increase their occupation.
\item $f(P>0.5)<0$. The orbitals that are initially occupied, upon increasing of $\lambda_s$ will steadily decrease their occupation.
\end{itemize}
As a result of the above conditions, and because the decrease in occupation of the occupied orbitals should be associated to an equal increase of the occupation of the unoccupied orbitals, the limiting case of a symmetric density of states leads to $f(0.5)=0$.

Let us now introduce Fermi-Dirac occupations. Because \eqn{ac1} is invariant upon shifting the orbital energies (the trace $\sum_i P_{ii}=0$), we can assume that orbital energies are referenced wrt the chemical potential (that is, $\mu$, the average of the potential felt by the electrons) and introducing another short-hand notation, we indicate $\Delta_i = \varepsilon_i - \mu$. Thus, the Fermi-Dirac occupations for a temperature $\tau$ become
\eqtn{ac5}{P_i^{s,\lambda_s=1}=\left[ e^{\frac{\Delta_i}{K_B\tau}} + 1 \right]^{-1},}
where $K_B$ is the Boltzman constant.

We propose two options for the interpolating function, $f(P)$:
\begin{enumerate}
\item We assume that $\lambda_s=\frac{T}{\tau}$ (where $0<T<\tau$) and that $P_i^{s,\lambda_s}=P=\left[ e^{\frac{\Delta_i}{K_B\tau}\left(\frac{1}{\lambda_s}\right)} + 1 \right]^{-1}$. By inverting and applying \eqn{ac7}, we find
\eqtn{ac8}{f(P)=\frac{K_B\tau}{\Delta_i}\frac{1}{P(1-P)}.}
The choice above, considering \eqn{ac1} and that $P^0=1$ if $\Delta_i<0$ and $P^0=0$ if $\Delta_i >0$, leads to the following expression for the static correlation energy,
\begin{align}
\nonumber
\label{ac9}
E_c^\text{static}=&K_B\tau ~\text{Tr}\big[ \text{ln}(1-P) \big], \text{when } P<0.5\\
 &+ K_B\tau ~\text{Tr}\big[ \text{ln}(P) \big], \text{when } P>0.5
\end{align}
where Tr indicates the trace operator. There is no need to define a contribution when $P=0.5$, because it implies that the canonical orbital energy is zero (i.e., equal to the chemical potential), that is $\Delta_i=0$.
\item Inspired by \eqn{th3}, the following option
 \eqtn{ac10}{f(P)=\frac{K_B\tau}{\Delta_i}\frac{1}{(P^0-P)}\text{ln}\left[ \frac{1-P}{P} \right],}
leads to 
\eqtn{ac11}{E_c^\text{static}=K_B\tau \text{Tr}\big[ P \text{ln}(P) + (1-P) \text{ln}(1-P) \big]=-\tau S_s^\tau}
\end{enumerate}

\begin{figure}
\includegraphics[width=0.8\textwidth]{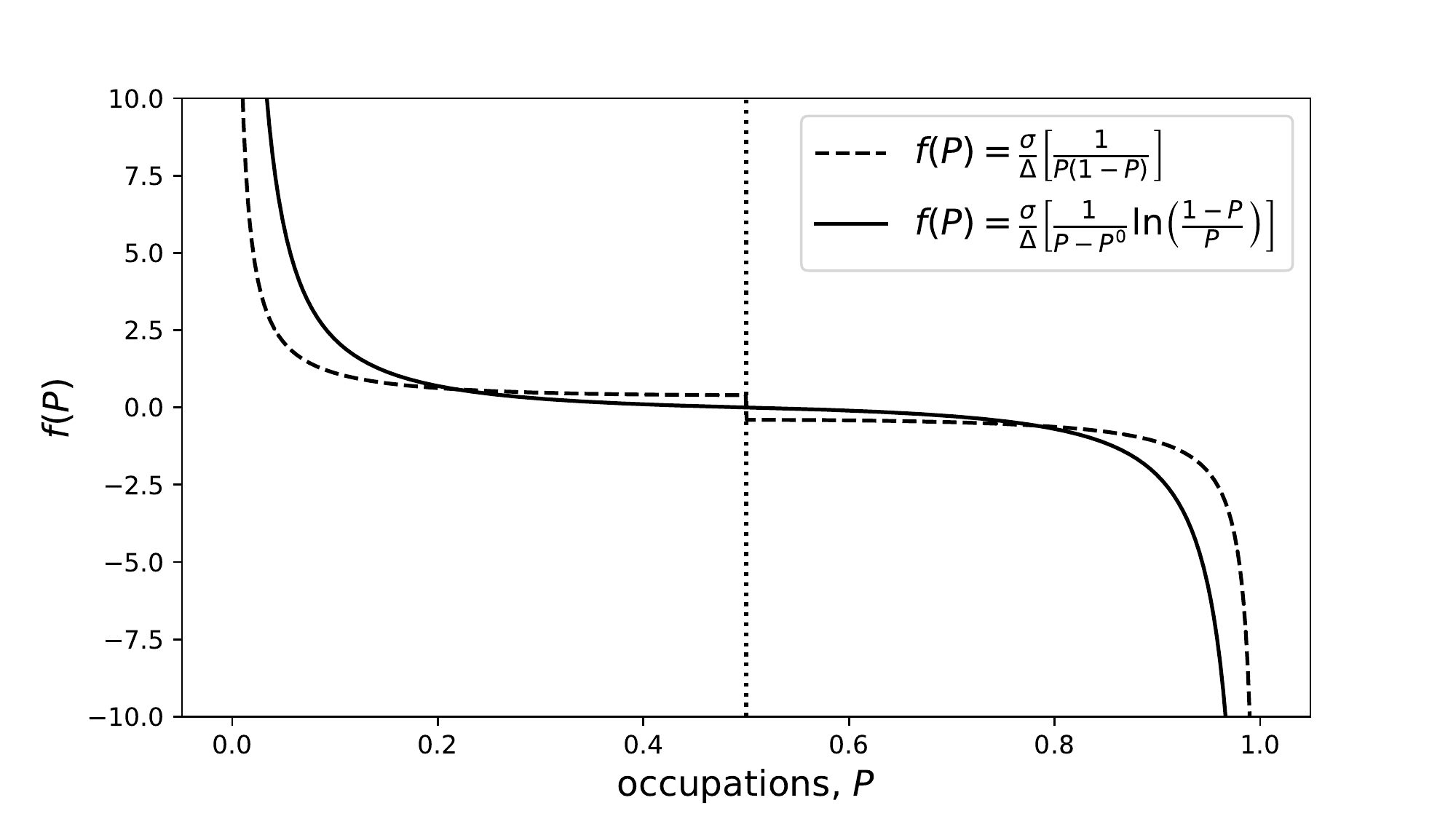}
\caption{\label{f_comp} Comparison of the two proposed options for $f(P)$ to be used in \eqn{ac6}. In the plot, we chose $\frac{K_B\tau}{\Delta}=0.1$.}
\end{figure}

A comparison of the two proposals for $f(P)$ is given in Figure \ref{f_comp}. From the figure, it is striking to note that the two options encode essentially the same physics. Thus, from a practical standpoint, we prefer the option that gives more accurate results in our pilot calculations presented in the following section. 

\section{Results and Discussion}
We present several proof of principle calculations involving bond stretching. We choose H$_2$ and LiH as examples of homodinuclear and heterodinuclear molecules. We also consider H$_2^+$ as a challenging system (arguably the worst case scenario) for a static correlation energy functional. In all cases, we compare the zero-temperature (or $K_B\tau=0$) KS reference with the thermal reference at $K_B\tau=1.0$ eV for H$_2$ and 0.3 eV for LiH and H$_2^+$. Calculations are carried out with PySCF \cite{PYSCF}. Unless otherwise specified, all calculations are carried out employing the {cc-pVTZ} basis set and the PBE exchange-correlation functional \cite{PBEc}.

\subsection{H$_2$: comparison of the two options for $f(P)$}

We must first establish whether Fermi-Dirac occupations are suitable for reproducing the canonical occupation numbers derived from the FCI 1-RDM. We compute the occupations by representing the FCI 1-RDM in the basis of the canonical Hartree-Fock orbitals. The DFT occupations are reported in the basis of the KS orbitals. All simulations are carried out to selfconsistency, also with respect to the particular choice of occupation numbers employed.

\begin{figure}
\includegraphics[width=1.0\textwidth]{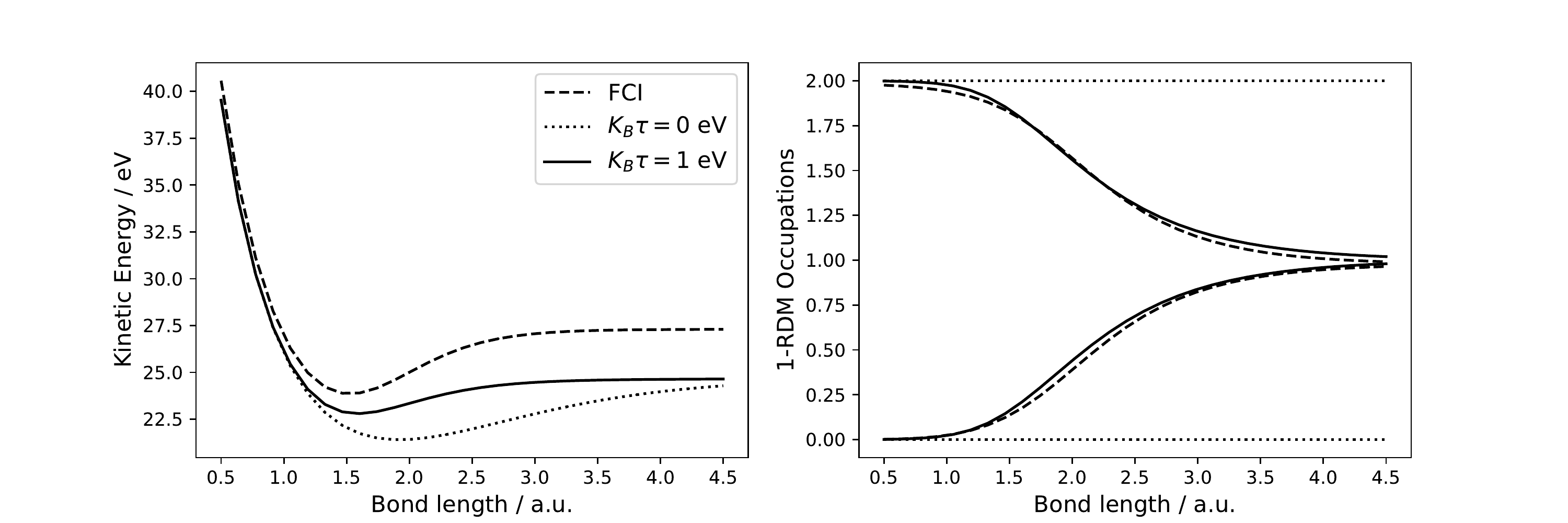}
\caption{\label{H2_1} H$_2$ dissociation curve. Left: total kinetic energy. Right: canonical occupation numbers.}
\end{figure}

From the right-hand-side of Figure \ref{H2_1}, we note a striking agreement between the canonical occupations of FCI and the thermal reference, suggesting that smearing the occupations, for example employing the Fermi-Dirac distribution, is a viable route to recovering static correlation via the adiabatic connection integral. Similar results were also reported by Chai \cite{Chai_2012}, however the FCI occupations were reported in terms of the FCI natural orbitals rather than the canonical HF orbitals. These results indicate that the Fermi-Dirac smearing can effectively accomplish the condition $P_i^{s,\lambda_s=1}$ close to $P_i^{\lambda_d=1}$.

From the left-hand side of Figure \ref{H2_1}, we evince that the kinetic energy of the thermal reference is closer to the interacting one in the region of internuclear separations that feature the Coulson-Fisher point. In this region, the character of the 1-RDM changes considerably as near degeneracy among the fronteer orbitals arise. In the asymptotic region, the noninteracting kinetic energies (whether thermal or not) converge to the same asymptote which is located below the one of FCI by about 2 eV. 

\begin{figure}
\includegraphics[width=1.0\textwidth]{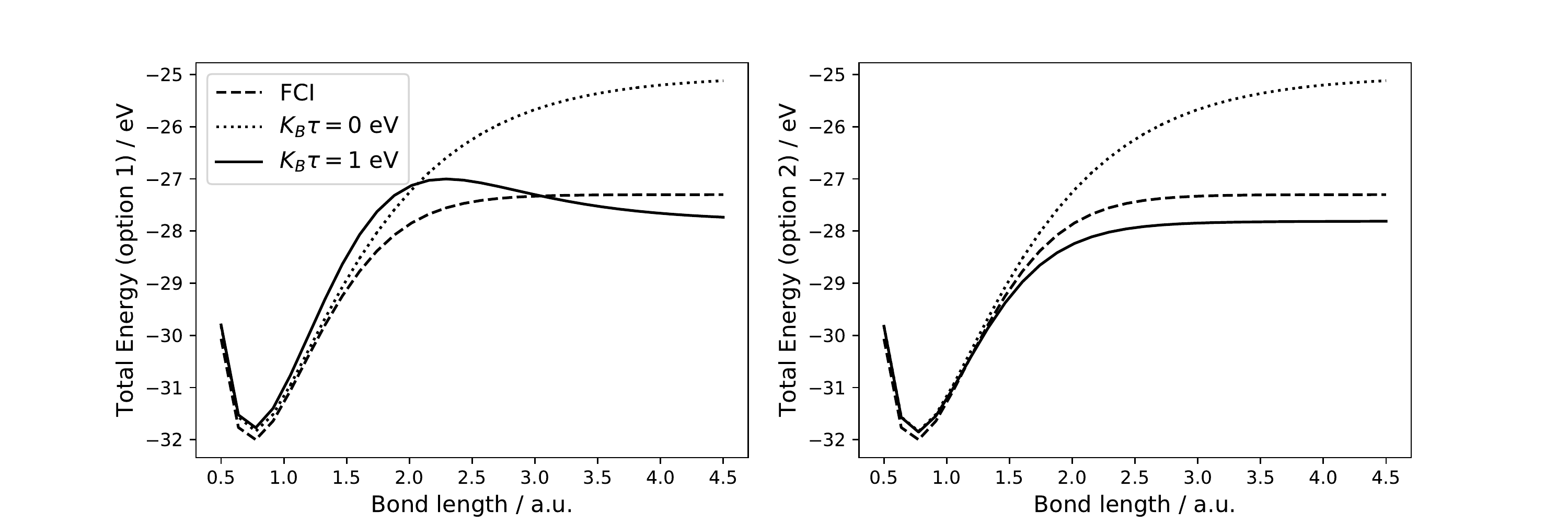}
\caption{\label{H2_2} H$_2$ dissociation curve. Total energy employing $E_c^\text{static}$ computed with \eqn{ac9} (option 1 on the lhs) or \eqn{ac11} (option 2 on the rhs) for $f(P)$. }
\end{figure}

In Figure \ref{H2_2}, we plot the potential energy curve for H$_2$ employing the two prescriptions for the static correlation energy functional, $E_c^\text{static}$, in \eqn{ac9} (option 1) and \eqn{ac11} (option 2). From the figure, we notice that although option 1 reduces the deviation from the FCI energy, the shape of the curve with this option is incorrect. Particularly, the hump at $R=2$ Bohr, is unphysical and so is approaching the dissociation asymptote from above.

Conversely, option 2 provides us with an improved energy curve resembling the FCI result. The thermal reference's dissociation asymptote deviates from FCI by about a half of 1 eV, compared to a deviation of more than 3 eV when the KS reference is used. However, a more accurate result could be obtained by tuning the fictitious temperature of the thermal reference state, $\tau$.

In the supplementary information document \cite{epaps}, we show equivalent results for the N$_2$ molecule employing the same effective temperature as we did for H$_2$.

\subsection{LiH dissociation}
The case of LiH is more complex than H$_2$. In LiH, the bond breaks homolithically, however the curve goes through an avoided crossing and changes character from ionic to covalent. Thus, we expect static correlation to play a role in the vicinity of the avoided crossing as well as in the asymptotic region.

\begin{figure}
\includegraphics[width=1.0\textwidth]{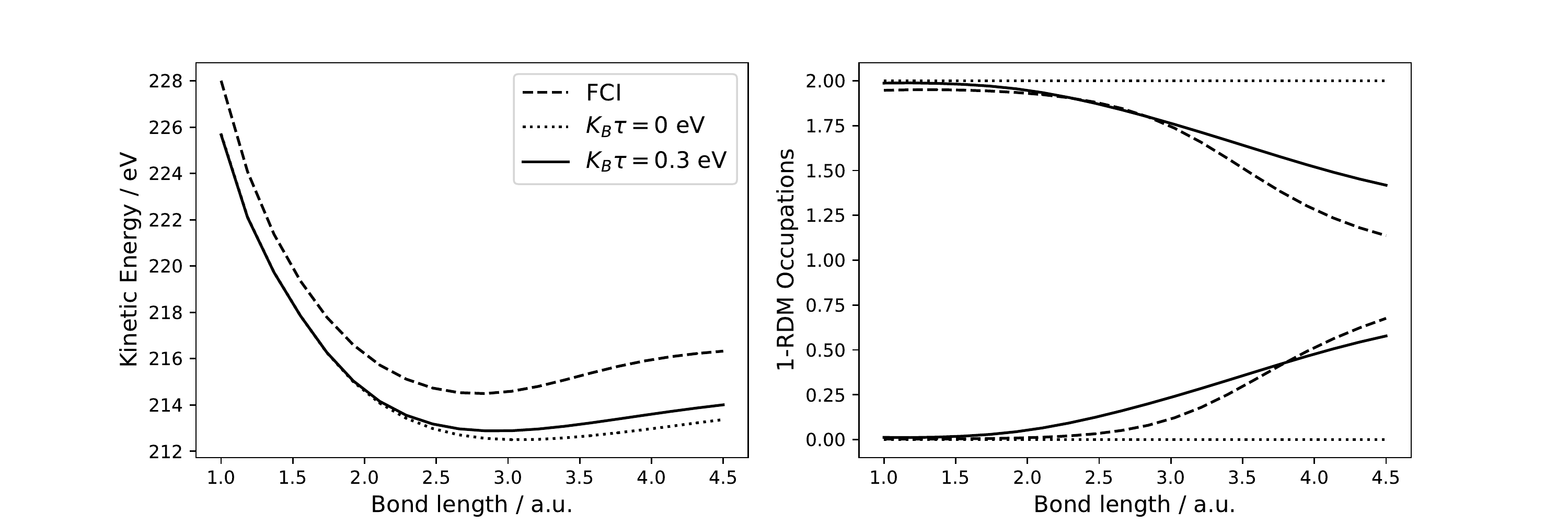}
\caption{\label{LiH_1} LiH dissociation curve. Left: total kinetic energy. Right: canonical occupation numbers. }
\end{figure}

Figure \ref{LiH_1} shows that LiH canonical occupations are reasonably well reproduced by the Fermi-Dirac distribution even though the effective temperature used is reduced from 1 eV for H$_2$ to 0.3 eV for LiH. 

\begin{figure}
\includegraphics[width=1.0\textwidth]{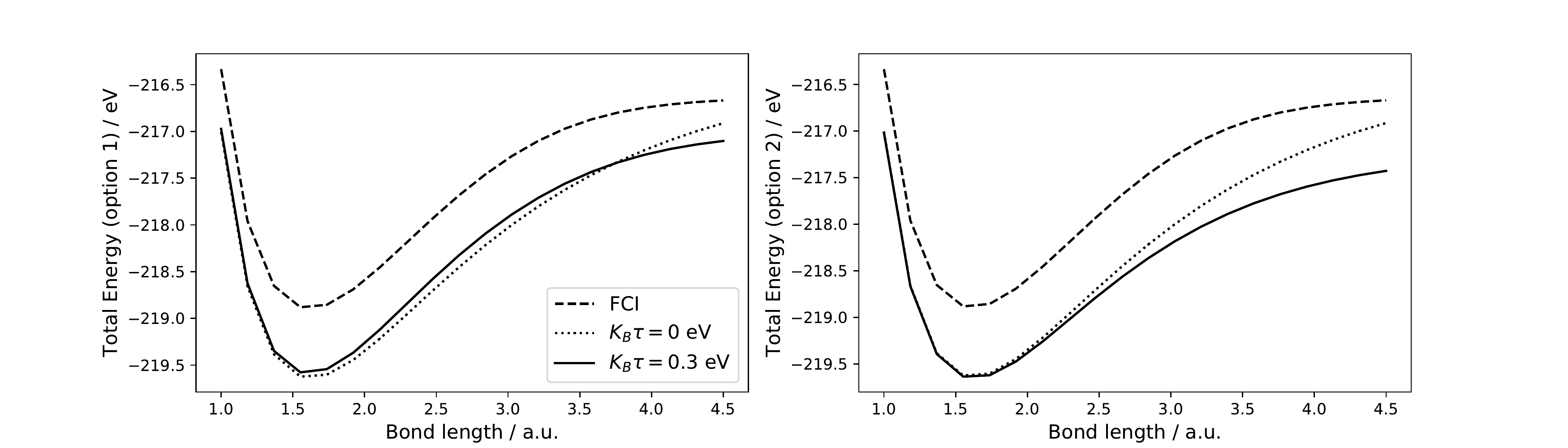}
\caption{\label{LiH_2} LiH dissociation curve. Total energy employing $E_c^\text{static}$ computed with \eqn{ac9} (option 1 on the lhs) or \eqn{ac11} (option 2 on the rhs) for $f(P)$. }
\end{figure}

Turning to the evaluation of the static correlation energy functional, comparing options 1 and 2 (with \eqn{ac9} and \eqn{ac11}, respectively) for the adiabatic connection path, we see in Figure \ref{LiH_2} that both options improve on the general asymptotics and shape of the energy curve. Option 2 leads to a smaller binding energy compared to option 1.

\begin{table}
\caption{\label{tab:LiH}LiH dissociation energy, $D_e$, in eV. $K_B\tau$ also in eV.}

\begin{tabular}{ccccc}
&&DFT& & ~~FCI \\ 
\cline{2-4}
$K_B\tau$ & $0$ & $0.3$ & $1.0$ & ~~-- \\ 
$D_e$ &2.27              & 2.23              & 1.07              & ~~2.23
\end{tabular}
\end{table}

Table \ref{tab:LiH} collects values of dissociation energies computed with FCI as well as KS DFT with thermal and pure KS states. As for the case of H$_2$, the results show that employing the static correlation functionals improves the dissociation provided that an appropriate value for the reference temperature, $\tau$ is used.

\begin{figure}
\includegraphics[width=0.7\textwidth]{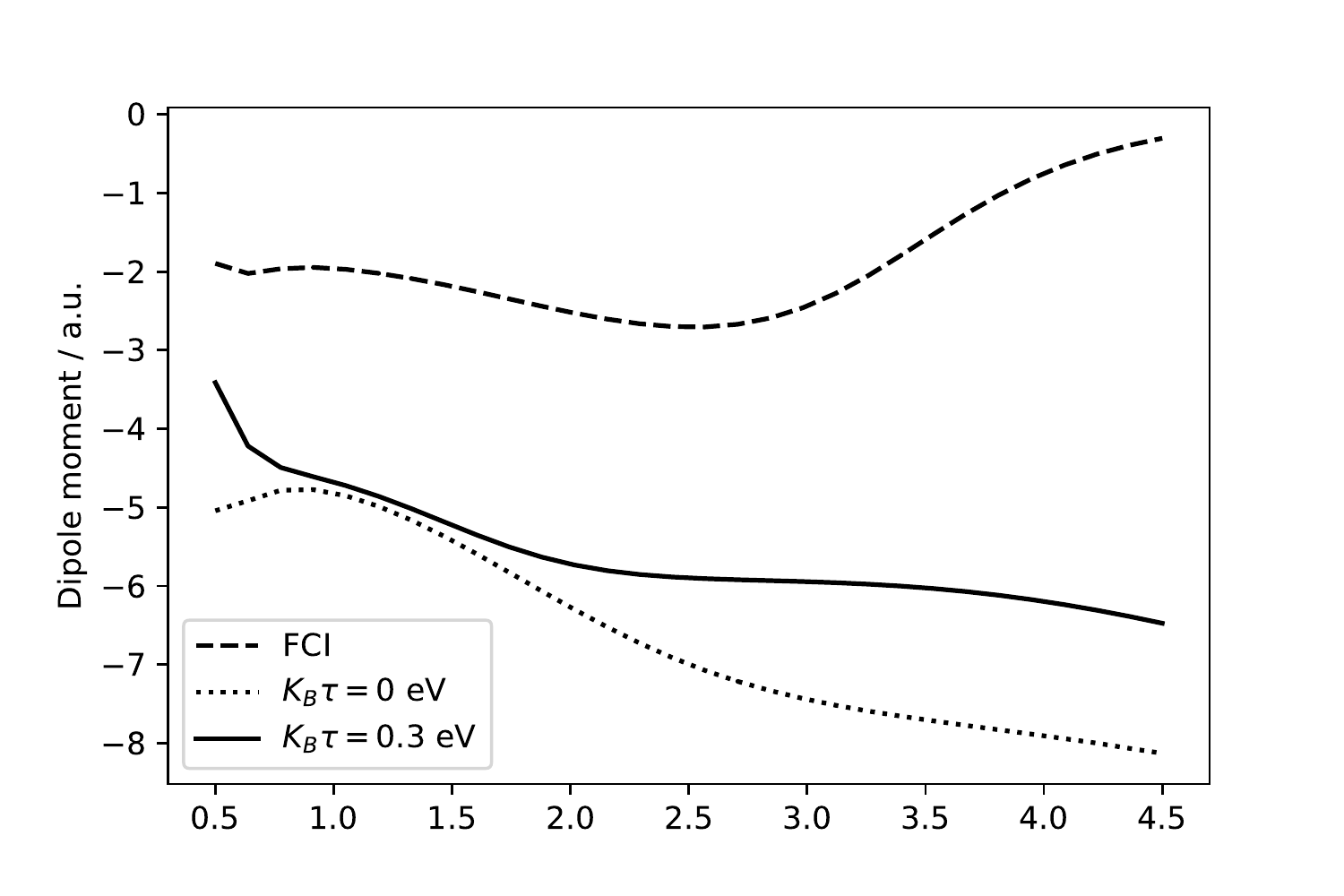}
\caption{\label{LiH_3} LiH dipole moment curve (in a.u.).  }
\end{figure}

One of the important features of using a thermal noninteracting state, is that the electron density may improve, especially where strong static correlation affects a system. Thus, to inspect this possible positive effect, we plot the dipole moment of LiH along the bond distance. See Figure \ref{LiH_3}.

Although the LiH dipole is much improved as a result of using a thermal noninteracting reference system, the improvement is not strong enough to recover the correct vanishing dipole moment in the asymptotic region. This is the result of a well-documented phenomenon: the (dis)appearance of step structures in the ground state KS potential \cite{Tempel_2009,Makmal_2011}. By allowing the KS density matrix to relax the occupations via Fermi-Dirac smearing, the detrimental effect due to the missing step structure in the KS potential are ameliorated. 

\subsection{H$_2^+$ dissociation}
H$_2^+$ is perhaps the most challenging system for any functional aiming at tackling static correlation \cite{Cohen_2011}. Because the static correlation functional is nonpositive by construction, it will deteriorate the energy of any system for which approximate xc functionals already yield too low energies. Perhaps the most physically meaningful explanation for the too-low energy of H$_2^+$ is due to Becke \cite{Becke_2003}, and it invokes the fact that local and semilocal functionals overestimate the exchange hole normalization (by construction it is imposed to be -1) in this system and miss the purely nonlocal shape of the true exchange hole (half of it sits around one proton and the other half around the other proton). 

Inspection of Figure S1 of the supplementary materials document \cite{epaps} shows that indeed there is no improvement for the bond-breaking energy curve of H$_2^+$, and following expectations \cite{Cohen_2011}, the addition of the static correlation functional slightly deteriorates the energy curve \cite{van_Aggelen_2010}. It is worth mentioning that the deterioration contributed by the addition of the static correlation functional is still much smaller than the existing deviation of semilocal KS-DFT from the FCI (HF in this case) dissociation asymptote.

\section{Conclusions}
We presented a mean-field framework for modeling electronic systems that feature strong static correlation. The formalism is based on an adiabatic connection approach that connects a reference noninteracting system to the correlated (real) system. The initial noninteracting reference is chosen to be an \emph{approximate} Kohn-Sham system computed with a single Slater determinant and an approximate exchange-correlation functional. The adiabatic connection path allows the electron density to vary. This is key, because in a static correlation regime the electron density computed with approximate functionals and single-reference wavefunctions are deviated from the exact case.

The theoretical framework allows us to develop static correlation density functionals as well as their dynamic correlation counterparts. For the static correlation part, a choice needs to be made in regards to a smearing procedure for the mean-field orbital occupations. In this work, we employed Fermi-Dirac occupations with an effective temperature, $\tau$, as a parameter of the method. However, other prescriptions are also possible \cite{Savin_2000}. We also showed that it is possible to define several adiabatic connection paths that associate the smearing procedure to a static correlation energy. Once the static correlation adiabatic connection path is chosen, the dynamic correlation can be evaluated. We show that current exchange-correlation approximants are viable choices for dynamic correlation functionals for systems with a gap.

The results of our computational investigations validate several previous observations that the electronic density is improved by smearing occupations, and that the noninteracting kinetic energy of a thermal state typically is closer to the interacting one compared to the one of a pure KS single Slater determinant. We further show that the electronic Shannon entropy is a good approximation for the static correlation density functional. Self-interaction in the xc density functional is still problematic, impeding the correct description of heterodinuclear molecules (such as LiH)  as well as the one-electron molecular ion, H$_2^+$.

Future investigations will involve generating a self-consistent procedure by taking (via OEP or direct differentiation in a Generalized KS scheme) the functional derivative wrt to the electron density of the static correlation density functionals. Additionally, we will continue exploring static correlation paths and smearing procedures beyond the Fermi-Dirac distribution. 

Other types of degenerate systems, such as atoms, will be the topic of future investigations especially in regards to the ability of the static correlation functionals to describe correctly an array of statically correlated systems beyond bond dissociation.
\begin{acknowledgments} 
This material is based upon work supported by the National Science Foundation under Grant No. CHE-1553993.
We are grateful to Andreas Savin, Neepa Maitra, Lionel Lacombe and Florian Eich for entartaining discussions on the topic and help for improving the presentation.

\end{acknowledgments}

\end{document}